\pdfoutput=1
\documentclass[final,5p,times,twocolumn,authoryear]{elsarticle}

\usepackage{amssymb}
\usepackage{amsmath}	
\usepackage{hyperref}
\usepackage{listings}
\usepackage{cellspace} %
\DeclareUnicodeCharacter{0393}{$\Gamma$} 
\DeclareUnicodeCharacter{2009}{ } 
\usepackage[dvipsnames, table]{xcolor}
\definecolor{codegreen}{rgb}{0,0.6,0}
\definecolor{codepurple}{rgb}{0.58,0,0.82}
\journal{Astronomy $\&$ Computing}

\begin{document}
	
	\begin{frontmatter}
		
		
		
		\title{\textsc{Magritte}, a modern software library for spectral line radiative transfer}
		
		
		\author[IvS]{T. Ceulemans}
		\author[IvS]{F. De Ceuster}
		\author[IvS]{L. Decin}
		\author[UCL]{J. Yates}

		\affiliation[IvS]{organization={Institute of Astronomy, Department of Physics and Astronomy},
			             addressline={Celestijnenlaan 200D},
			             city={Leuven},
			             postcode={3001},
			             country={Belgium}}
		\affiliation[UCL]{organization={Department of Computer Science, University College London},
		addressline={Gower Place},
		city={London},
		postcode={WC1E 6BT},
		country={UK}}		             
		
		\begin{abstract}
        Spectral line observations are an indispensable tool to remotely probe the physical and chemical conditions throughout the universe.
        Modelling their behaviour is a computational challenge that requires dedicated software.
        In this paper, we present the first long-term stable release of \textsc{Magritte}, an open-source software library for line radiative transfer.
        First, we establish its necessity with two applications.
        Then, we introduce the overall design strategy and the application/programmer interface (API).
        Finally, we present three key improvements over previous versions:
        (1) an improved re-meshing algorithm to efficiently coarsen the spatial discretisation of a model;
        (2) a variation on Ng-acceleration, a popular acceleration-of-convergence method for non-LTE line transfer; and,
        (3) a semi-analytic approximation for line optical depths in the presence of large velocity gradients.
		\end{abstract}
		
		
		

	\end{frontmatter}
	
	
	

\section{Introduction}
\label{section: Introduction}
Electromagnetic radiation plays a key role in astrophysics, as it governs several crucial physical and chemical processes.
For instance, radiation pressure can shape the dynamics of stellar atmospheres \citep[see e.g.][]{goldreich_oh-ir_1976, castor_radiation_2004, moens_first_2022}, radiative heating and cooling can significantly impact energy balance \citep[see e.g.][]{vernazza_structure_1981, neufeld_thermal_1995, esseldeurs_3d_2023}, and various photo-reactions can have a considerable impact on chemistry \citep[see e.g.][]{woitke_radiation_2009, van_de_sande_chemical_2020}.
Moreover, radiation transport dictates what we can and cannot observe.

Electromagnetic radiation contains a wealth of information,
in particular, spectral lines.
These sharp features in frequency space originate from transitions between the quantized energy levels of electronic, rotational, or vibrational states of the atoms and molecules in the medium.
Their specific frequencies uniquely characterise the chemical species that produces them and thus can be used to infer the chemical composition of the medium.
Spectral lines also encode the kinematic structure of the environment.
Their narrow extent in frequency space makes them very sensitive to Doppler shifts caused by macroscopic velocity gradients in the medium, while their widths are governed by microscopic thermal and turbulent motions.
Furthermore, the relative intensities of lines can be used to infer the thermodynamic properties of the environment.
As such, spectral line are excellent tracers of both the physical and chemical properties of the medium from which they originate, making them indispensable for astrophysics research.
In order to leverage all this information, over the years, several software models have been developed to simulate line radiative transfer, such as
\textsc{MCFOST} \citep{pinte_monte_2006}, \textsc{LIME} \citep{brinch_lime_2010}, \textsc{SKIRT} \citep{baes_skirt_2015, matsumoto_self-consistent_2023}, and also \textsc{Magritte} \citep{deceuster_magritte_2020, deceuster_magritte_2020-1, de_ceuster_3d_2022}.

In this paper, we present the first stable release of \textsc{Magritte}, a modern software library for spectral line radiative transfer.
Previous papers about \textsc{Magritte} have focused on the specific algorithms for ray-tracing, the self-consistent solution of the radiation field with the state of the medium, and benchmarks against other codes \citep{deceuster_magritte_2020}, and also on reduction methods for the model discretisation and their effect on the computational efficiency \citep{deceuster_magritte_2020-1}.
In this paper, we focus on the now-consolidated design strategy with the now-stable application/programmer interface (API), which mark the corresponding long-term stable release of the source code, \footnote{Publicly available at \url{github.com/Magritte-code/Magritte}.} while also presenting several further algorithmic improvements over previous versions \citep{de_ceuster_3d_2022}.

This paper is organized as follows.
In Section \ref{section: statement of need}, we demonstrate the utility of \textsc{Magritte} with two important applications.
In Section \ref{section: data structure algorithm design}, we describe the code structure and elaborate on our software development and maintenance strategy.
We present three key improvements in \textsc{Magritte} in Section \ref{section: computational considerations}, and we discuss current limitations and future work in Section \ref{sec:limitations}.

\section{Statement of need}
\label{section: statement of need}
The era of spatial and spectral high-resolution imaging, such as with the Atacama Large (sub)Millimetre Array (ALMA) or the Square Kilometre Array (SKA), has pushed the need for versatile radiative transfer simulators that can handle high-resolution models to simulate all complex phenomena that are observed \citep[see e.g.][]{andrews_disk_2018, decin_substellar_2020, oberg_molecules_2021}.
\textsc{Magritte} is a radiative transfer simulator that was primarily designed with applications in mind of modelling stellar environments.
Since its conception, it has been used, for instance, to model the unusual NaCl distribution around asymptotic giant branch (AGB) star IK Tauri \citep{coenegrachts_unusual_2023-1}, to model the molecule-rich disk around AGB star L$_{2}$ Puppis \citep{vandesande_modelling_2024}, and to create synthetic observations of smoothed-particle hydrodynamics (SPH) models of companion-perturbed stellar winds (Malfait et al.\ subm.).
Furthermore, \cite{esseldeurs_3d_2023} implemented the ray-tracer of \textsc{Magritte} in the SPH code \textsc{Phantom} \citep{price_phantom_2018} to couple radiation and hydrodynamics.
Below, we demonstrate how \textsc{Magritte} can be used to model the CO line emission of an analytic model of a protoplanetary disk and an intricate SPH model of a companion-perturbed stellar wind.

\begin{figure}
	\includegraphics[width=\linewidth]{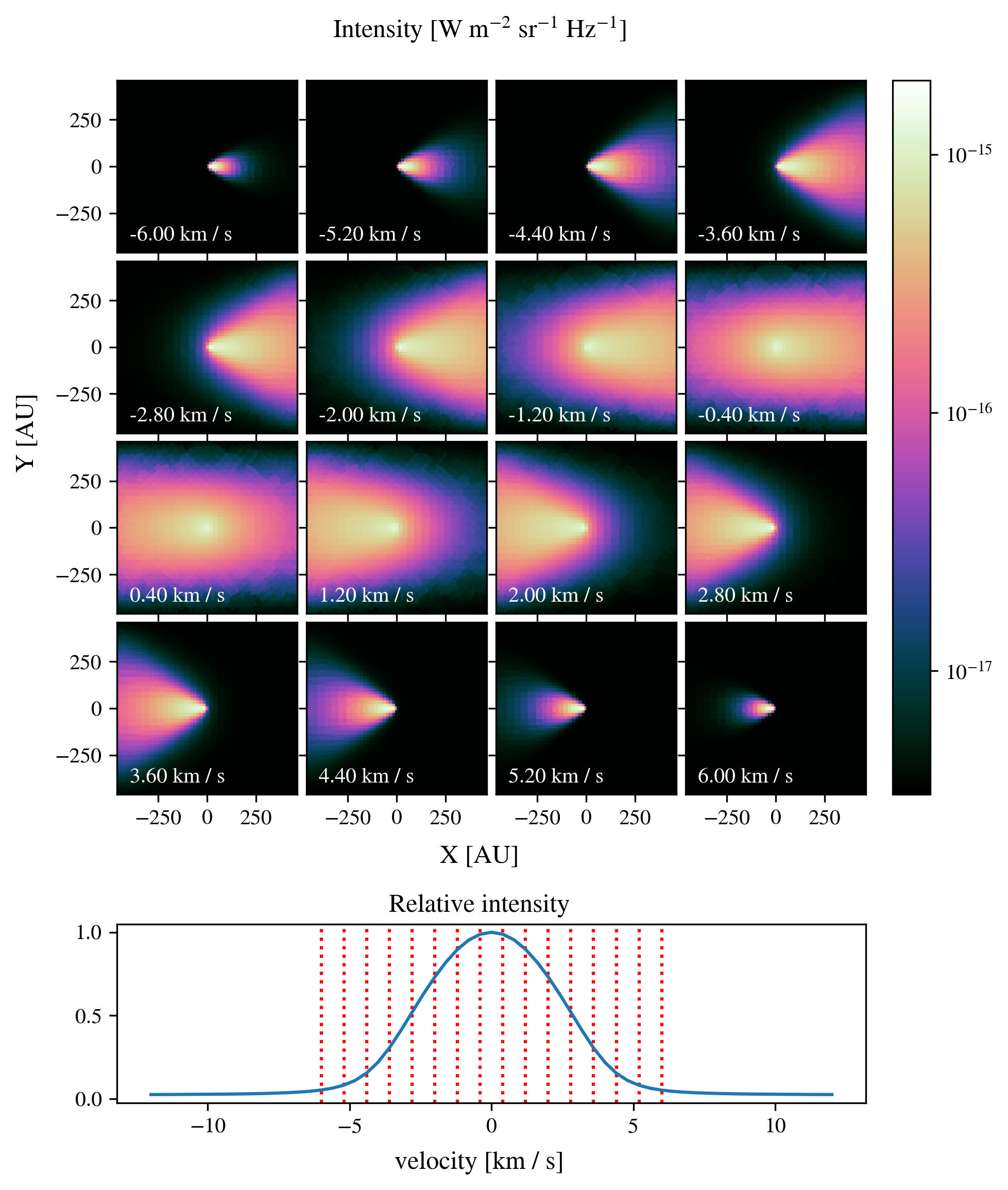}
	\caption{Synthetic observations of the CO $J=1-0$ line of an analytic protoplanetary disk model. \textit{(Top.)} Specific intensity in each frequency/velocity bin. \textit{(Bottom.)} Normalized spatially integrated intensity.}\label{fig: channel maps analytic disk}
\end{figure}

\begin{figure}
	\includegraphics[width=\linewidth]{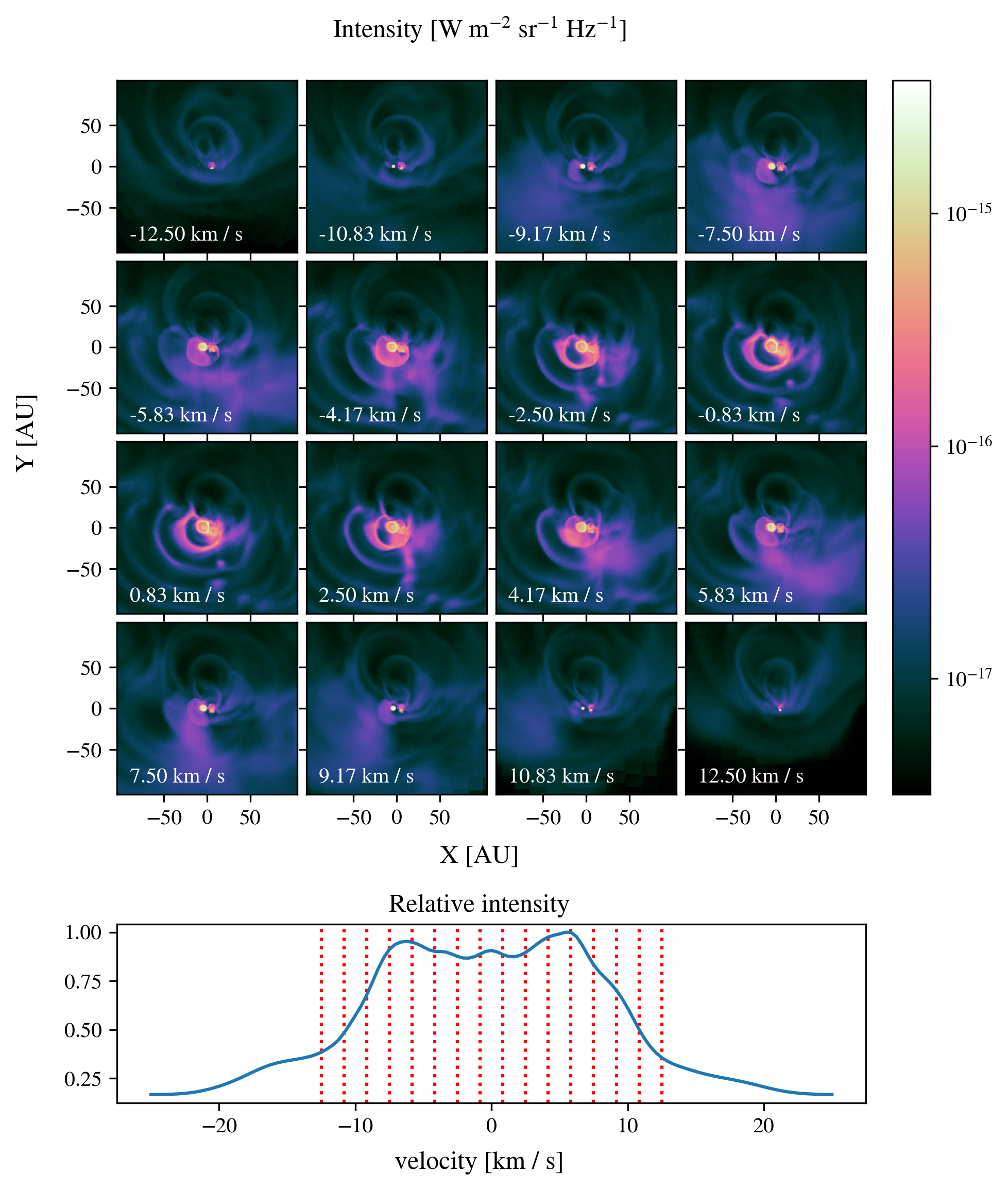}
	\caption{Synthetic observations of the CO $J=2-1$ line of an SPH simulation of a companion-perturbed stellar wind model. \textit{(Top.)} Specific intensity in each frequency/velocity bin. \textit{(Bottom.)} Normalized spatially integrated intensity.}\label{fig: channel maps phantom original mesh}
\end{figure}

\subsection{Analytic protoplanetary disk model}
\label{section: protoplanetary disk}
Figure \ref{fig: channel maps analytic disk} shows synthetic observations created with \textsc{Magritte} of the CO $J=1-0$ line emission emanating from a proto-planetary disk model, viewed edge-on.
A detailed description of the model can be found in \citep{deceuster_magritte_2020}\footnote{\cite{deceuster_magritte_2020} also showed synthetic observations of this model. However, due to boundary effects, the integrated line profile was inaccurate. This has now been resolved. The computational improvements presented in this paper, moreover, allow us now to create much higher resolution images.}.
The \textsc{Magritte} model used a discretisation of about 100k points, 48 directions, and 7 frequency bins per line.
The radiative properties of the medium\footnote{This means the emissivities/opacities computed using the populations of the 41 lowest rotational energy levels of CO. As for every model in this paper, the line data were obtained from the LAMDA database, which can be found at \url{home.strw.leidenuniv.nl/~moldata/}.} were determined self-consistently with the radiation field, without assuming local thermodynamic equilibrium (i.e.\ non-LTE).
The corresponding channel maps are shown in Figure \ref{fig: channel maps analytic disk}.
These show the intensity in different frequency bins, which, through the Doppler shift, can be related to the emission of material moving with a specific velocity along the line of sight.
When scanning through the different channels from top-left to bottom-right, we can see the emission shift from right to left with respect to the centre of each image.
This is a clear signature of the rotation of the disk.
The detailed structure of the line emission of the disk can further be used to infer physical and chemical properties, such as the velocity and temperature distribution \citep[see e.g.][]{vandesande_modelling_2024}.

\subsection{Numerical companion-perturbed stellar wind model}
\label{section: AGB binary}
\begin{figure}
	\centering
	\includegraphics[width=\linewidth]{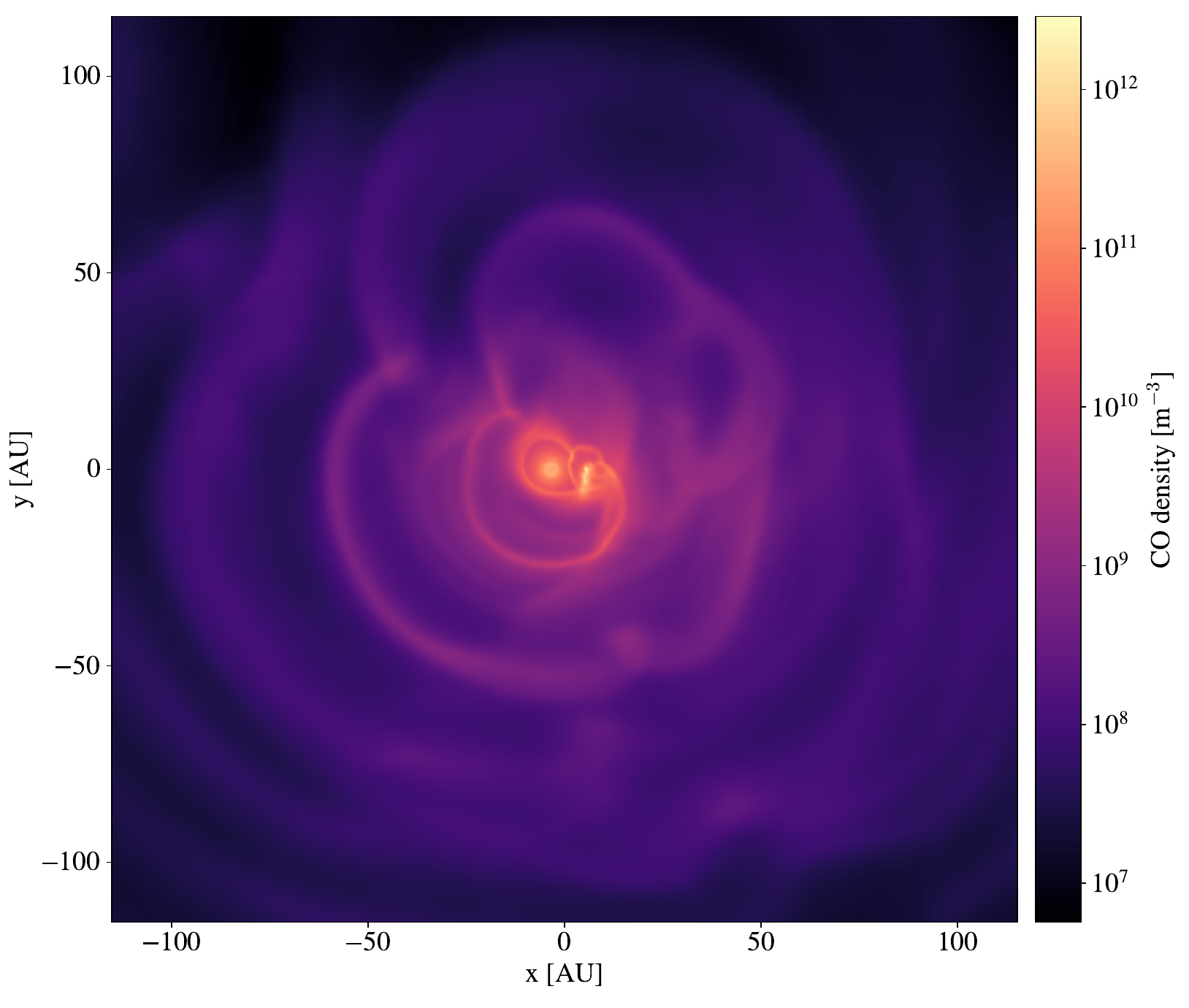}
	\caption{Slice through the centre of the companion-perturbed stellar wind model showing the CO number density.}
 \label{fig: AGB model density}
\end{figure}

Figure \ref{fig: channel maps phantom original mesh} shows synthetic observations created with \textsc{Magritte} of the CO $J=2-1$ line emission emanating from an SPH simulation of a companion-perturbed stellar wind.
The line-of-sight is orthogonal to the orbital plane of the companion.
A star in the centre is losing mass in an initially spherically-symmetric wind, stirred up by a companion on an eccentric orbit.
Figure \ref{fig: AGB model density} depicts a slice through the centre of the model, showing the intricate wind structure in terms of the CO number density.\footnote{To obtain the CO number density from the SPH model (that only contains the total mass density), we attribute all the mass density to H$_{2}$ and then assume a constant ratio of $10^{-4}$ between the CO and H$_{2}$ number densities.}
A detailed description of the SPH model can be found in \cite{malfait_sph_2021}.
The SPH model, as processed by \textsc{Magritte}, contains about 1M points, with 48 directions and 7 frequency bins per line to discretise the radiation field.
Using \textsc{Magritte}, first, non-LTE line transfer is performed to determine the level populations of the CO rotational levels, and then the images are created.
Given the complexity of the model, the interpretation of the synthetic observations is less straightforward as with the previous disk model.
The channel maps in Figure \ref{fig: channel maps phantom original mesh} clearly show the arcs that are stirred up by the companion in the stellar wind.
In the higher-velocity images we can even discern the mass-losing star and its companion through the hot and fast-moving material that enshrouds them.
The velocity structure is quite symmetric, due to the initially spherically-symmetric velocity field of the wind and the fact that the line-of-sight is orthogonal to the orbital plane.

\section{\textsc{Magritte} design}
\label{section: data structure algorithm design}
The design of \textsc{Magritte} is mainly driven by the need for a modular and performant general-purpose radiation transport simulator that can easily be applied to grid-based as well as particle-based model data.
The resulting design choices for the geometric data structures, the model discretisation, and the ray-tracing algorithm have extensively been discussed in \cite{deceuster_magritte_2020, deceuster_magritte_2020-1} and \cite{de_ceuster_simulating_2022}.
Here, we focus on the code structure, the application / programmer interface (API), and the development and maintenance strategy.

\subsection{Code structure}
The core of \textsc{Magritte} is written in \textsc{C++}, but it has a convenient user-interface in \textsc{Python} that is linked to the \textsc{C++} core using \textsc{pybind11}.\footnote{Publicly available at \url{github.com/pybind/pybind11}.}
The \textsc{Magritte} core is written in an object-oriented fashion, organising model data in a physically relevant class structure.
The main class structure of a \textsc{Magritte} model is given in Listing \ref{listing: class}.
The \textsc{Parameters} class stores the general model parameters, such as a name for the model, the number of points, directions, and frequency bins, as well as algorithmic parameters, such as various threshold values.
Since all classes require this data, each model subclass has access to the \textsc{Parameters} class through a \textsc{C++} smart pointer (\verb|std::shared_ptr|).
The contents of the other  classes should be evident from their names and is further illustrated in the exposition of the API.

\begin{lstlisting}[caption={Model class structure in \textsc{Magritte}.}, label={listing: class}]
Model.Parameters

Model.Geometry
Model.Geometry.Boundary
Model.Geometry.Points
Model.Geometry.Rays

Model.Lines
Model.Lines.LineProducingSpecies

Model.Thermodynamics
Model.Thermodynamics.Temperature
Model.Thermodynamics.Turbulence

Model.Radiation
Model.Radiation.Frequencies

Model.Chemistry
Model.Chemistry.Species
\end{lstlisting}

\subsection{Application / programmer interface (API)}
\label{section: magritte model}
We introduce the \textsc{Python} API of \textsc{Magritte} with an example of how to set up a \textsc{Magritte} model. A comprehensive reference of the API together with several example use cases can be found in the online documentation.\footnote{\label{docs}Publicly available at \url{magritte.readthedocs.io}.}

Listing \ref{listing: Magritte model setup} shows a typical \textsc{Python} script that can be used to create a \textsc{Magritte} model.
We start, on lines 1-3, with importing \textsc{numpy} \citep{harris_array_2020} to create data structures that we can bind to \textsc{C++}, and importing the relevant \textsc{Magritte} modules.
Then, on line 4, we create a \textsc{Magritte} \textsc{Model} object.
On lines 7-14, we define model parameters, such as model name (line 7), number spatial dimensions in the model (line 8), number of points (here stored in a variable, \textsc{npoints}; line 9), number of rays per point (line 10), number of chemical species (line 11), number of line-producing species (line 12), number of Gauss-Hermite quadrature points that are used to integrate line profile functions (line 13), and number of boundary points (here derived from the length of the \textsc{numpy} array, \textsc{bdy\_npy}, which is assumed to contain the indices of the points that lie on the model boundary).
Next, in lines 16-20, the geometry is set, with, in particular, the position (line 16) and velocity (line 17) vector arrays. Both are assumed to be \textsc{numpy} arrays of shape (\textsc{npoints}, 3), since we assumed a 3-dimensional model. The nearest neighbours of each point are stored in a linearised array, \textsc{nbs\_npy} (line 18). Since the array is linearised, we also need the number of nearest neighbours for each point, \textsc{n\_nbs\_npy} (line 19), to be able to unpack the data.
The indices of the boundary points, \textsc{bdy\_npy}, are set on line 20.
On lines 22-23, the gas temperature and the (square of the) microturbulent velocity are set by \textsc{numpy} arrays that specify these values for each point.
Chemistry is defined, on lines 25-26, by a \textsc{numpy} array with the abundances for each species at each point, as well as the names of the species (in order of appearance in the abundance array).
In lines 28-31, we invoke several setup scripts (defined in \textsc{magritte.setup}) that define: a uniform distirbution of rays for each point (line 28), incoming cosmic microwave background (CMB) radiation on the boundary of the model (line 29), the LAMDA line data file \citep[][line 30]{schoier_atomic_2005}, and set up the Gauss-Hermite quadrature (line 31).
Finally, on line 33, we write the model to disk.
If no file name is given, the model is stored in a file with the model name (defined on line 7).

\begin{lstlisting}[caption={Example \textsc{Python} script for creating a \textsc{Magritte} model.}, label={listing: Magritte model setup}, language=python, numbers=left]
import numpy as np
import magritte.setup as setup
import magritte.core as magritte

m = magritte.Model()

m.parameters.set_model_name(name)   
m.parameters.set_dimension(3)            
m.parameters.set_npoints(npoints)      
m.parameters.set_nrays(48)            
m.parameters.set_nspecs(2)
m.parameters.set_nlspecs(1)
m.parameters.set_nquads(7)
m.parameters.set_nboundary(len(bdy_npy))

m.geometry.points.position.set(position_npy)
m.geometry.points.velocity.set(velocity_npy)
m.geometry.points.neighbors.set(nbs_npy)
m.geometry.points.n_neighbors.set(n_nbs_npy)
m.geometry.boundary.boundary2point.set(bdy_npy)

m.thermodynamics.temperature.gas.set(tmp_npy)
m.thermodynamics.turbulence.vturb2.set(trb2_npy)

m.chemistry.species.abundance = abundances_npy
m.chemistry.species.symbol    = ['CO', 'H2']

setup.set_uniform_rays(m)
setup.set_boundary_condition_CMB(m)
setup.set_linedata_from_LAMDA_file(m, lamda_file)
setup.set_quadrature(m)

m.write()
\end{lstlisting}

Once the model is created, we can simulate the radiation transport.
Listing \ref{listing: Magritte RT} shows a typical \textsc{Python} script to perform a radiative transfer simulation with \textsc{Magritte}.
In lines 1-3, we initialise the model, setting the frequency discretisation, computing the inverse line width, and initialising the level populations with their LTE values.
On line 5, we invoke the non-LTE solver for the level populations, which will iteratively and alternately compute the radiation field and the state of the medium \citep[see][for more details]{deceuster_magritte_2020}.
The first parameter, \textsc{Ng\_acc}, is a boolean that indicates whether or not to use the Ng acceleration of convergence method \citep[][see also Section \ref{section: ng-acceleration}]{ng_hypernetted_1974, olson_rapidly_1986-1}. The second parameter indicates the maximum number of iterations.
Once the state of the medium is determined, we can create synthetic observations of the model.
First, on line 7, we define the frequency range for the observations by its minimum (\textsc{f\_min}), its maximum (\textsc{f\_max}), and the number of frequency bins (\textsc{N\_freq}).
Finally, on line 9, we can create an image, along the ray direction, given by its components: \textsc{ray\_x}, \textsc{ray\_y}, and \textsc{ray\_z}.
The number of pixels along the image axis are given by \textsc{N\_pix\_x} and \textsc{N\_pix\_y}.
We refer to the documentation for a complete list of options.

\begin{lstlisting}[caption={Example \textsc{Python} script for radiative transfer in \textsc{Magritte}.}, language=python, label={listing: Magritte RT}, keywords={}, numbers=left]
m.compute_spectral_discretisation()
m.compute_inverse_line_widths()
m.compute_LTE_level_populations()

m.compute_level_populations_sparse(
    Ng_acc, N_iter_max)
m.compute_spectral_discretisation(
    f_min, f_max, N_freq)
m.compute_image_new(
    ray_x, ray_y, ray_z, N_pix_x, N_pix_y)
\end{lstlisting}

\textsc{Magritte} also provides convenient functionality to visualise and store the results.
Listing \ref{listing: Magritte tools} shows how the last synthetic observations that were created for a model can be visualised using \textsc{matplotlib} \citep[][line 4]{hunter_matplotlib_2007}, and also can be stored as a FITS file (line 5).

\begin{lstlisting}[caption={Example \textsc{Python} script for processing a \textsc{Magritte} image.}, language=python, label={listing: Magritte tools}, numbers=left]
import magritte.plot as plot
import magritte.tools as tools

plot.image_mpl(m)
tools.save_fits(m)
\end{lstlisting}

\subsection{Code development \& maintenance strategy}
\label{section: code maintenance}
\textsc{Magritte} is a collaborative project that has multiple developers and several contributors.
To facilitate the collaborative development process, we adopted the following strategy.

\subsubsection{Version control}
In \textsc{Magritte}, we use \textsc{git} for version control.
The repository is hosted online on \textsc{GitHub}.
Different projects and features are developed in different branches.
The latest stable version, the one that is recommended for scientific use, can be found on the \textsc{stable} branch.

To facilitate communication regarding updates, an automated versioning scheme has been implemented. This means that whenever the \textsc{stable} branch is updated, the version number of the source code will increase automatically. This provides users with a consistent and human-readable way to determine code versions, in addition to the less human-readable git hashes.

\subsubsection{Automated testing}
\label{section: automated testing design}
In order to swiftly catch the unavoidable errors that will be introduced by continued development on an existing code base, we introduced an automated testing framework in \textsc{Magritte}.
We use \textsc{pytest} for defining the test cases and automate them in our online repository using \textsc{GitHub Actions}.
We currently do not have unit tests but only regression tests.
The tests compare results obtained with \textsc{Magritte} against analytical results when possible, and, otherwise, with previously-obtained and validated numerical results.
The analytical tests that we now automated have previously been described in \cite{deceuster_magritte_2020}.
To test the non-LTE solver, we use the benchmarks described in \cite{van_zadelhoff_numerical_2002}.
We also include one ``real-world'' scientific application in our test suite, namely, the imaging of the SPH model presented in Section \ref{section: AGB binary}.

We support \textsc{Magritte} both on Linux and MacOS and thus also test on both operating systems.\footnote{Windows users are encouraged to use \textsc{Magritte} through the Windows Subsystem for Linux (WSL).}
We noticed that minor differences can be found between the online numerical test results on the two operating systems, mainly because the MacOS version is executed on ARM M1 Apple silicon, on which the \textsc{C++ long double} type has only 64 bits, while it has 80 bits on the \textsc{x86} hardware on which the Linux tests are executed.
Despite the numerical difference, both solutions safely reside within the assumed error margins.

\section{New features \& Improvements}
\label{section: computational considerations}
Since the last publication about \textsc{Magritte} \citep{de_ceuster_3d_2022}, some new features and several improvements have been introduced.
We highlight the three most important ones below.

\subsection{Recursive re-meshing}
\label{section: remeshing}
As in many numerical methods, also in radiation transport, discretisation plays a key role in the trade-off between accuracy and computational speed.
The computational cost of a ray-tracing radiative transfer solver roughly scales as $N^{4/3}$, with, $N$, the number of points in the spatial discretisation, while it scales linearly with respect to other parameters, such as the directional and frequency discretisation.
Therefore, it is crucial to be able to optimally choose the spatial discretisation.
\cite{deceuster_magritte_2020-1} proposed a simple algorithm to tailor the spatial discretisation of a hydrodynamics model to radiative transfer simulations.
The algorithm coarsened the spatial discretisation, based on a criterion on the gradient of a so-called tracer function, i.e.\ the function we want to properly sample with the (reduced) discretisation.
This was implemented in \textsc{Magritte}, using the mesh-generation functionality provided by \textsc{gmsh}\footnote{Publicly available at: \url{gmsh.info/}.} \citep{geuzaine_gmsh_2009} and \textsc{scipy} \citep{virtanen_scipy_2020}.
Although effective, generating new discretisations was computationally expensive \citep[it was e.g.\ a limiting factor in the work by][]{coenegrachts_unusual_2023-1}, and the implementation was rather cumbersome.
To alleviate this, we have implemented a much faster and simpler recursive re-meshing algorithm that we describe below.

\begin{figure}
	\includegraphics[width=\linewidth]{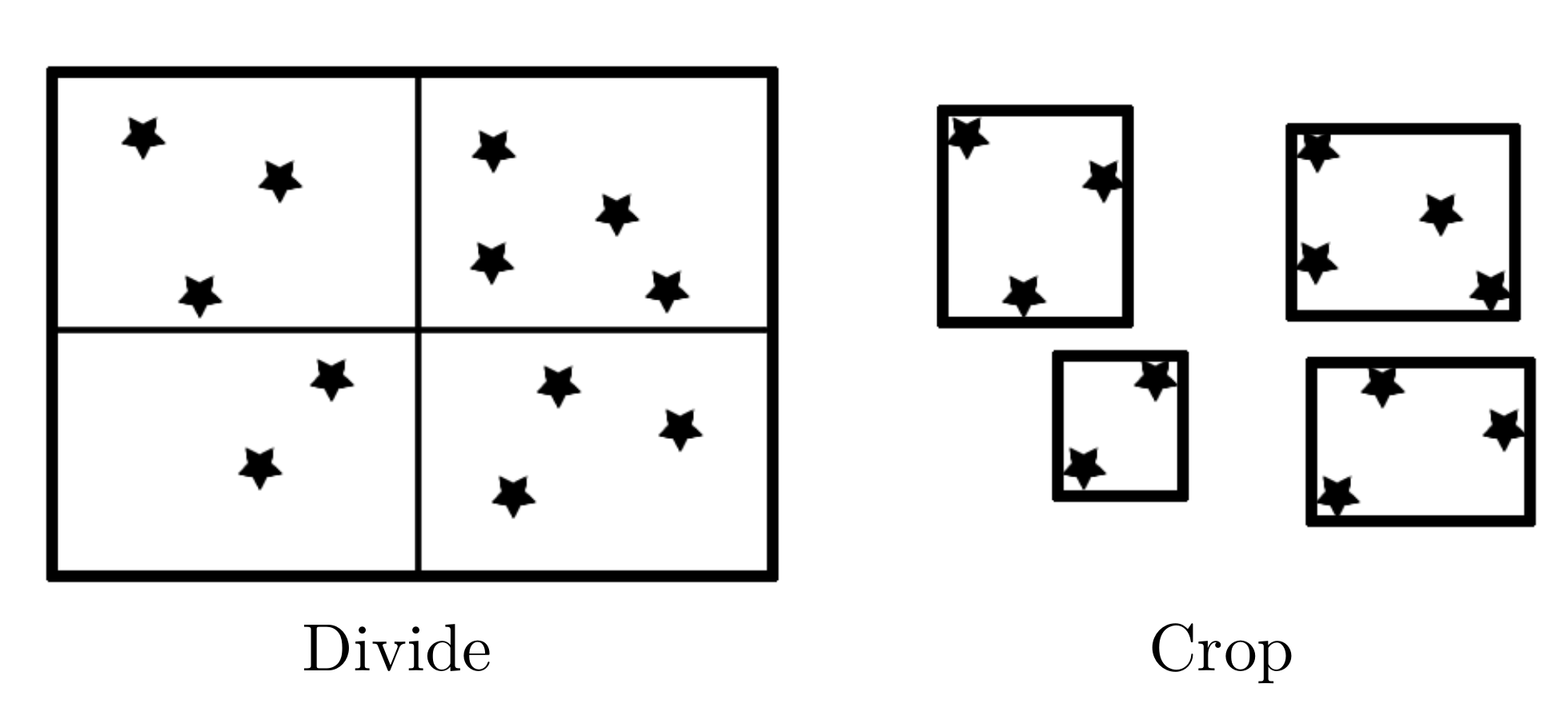}
	\caption{Illustration of the ``divide'' and ``crop'' steps in the recursive re-meshing algorithm. The stars denote the data points of the point cloud.}\label{fig: re-meshing procedure}
\end{figure}

Following \cite{deceuster_magritte_2020-1}, we assume the density, $\rho$, of a model to be an appropriate tracer function, i.e.\ we search for a coarser spatial discretisation that still properly samples the density.
Note, however, that we can replace this function by any other (combination of) model function(s), if needed.
In accordance with the internal data structures in \textsc{Magritte}, we assume the density to be given on a point cloud.
The recursive re-meshing algorithm then goes as follows.
Consider a rectangular box containing all data points and compute the variation of the tracer function within that box.
We define the variation within a box as
\begin{equation}\label{eq: subdivision criterion}
2 \left( \frac{\rho_{\max} - \rho_{\min}}{\rho_{\max} + \rho_{\min}} \right),
\end{equation}
where $\rho_{\max}$ and $\rho_{\min}$ are respectively the maximum and minimum value of the tracer function attained within the box.
If this variation exceeds a certain threshold value, $r_{\max}$, then we deem the variation too large and subdivide the rectangular box in 4 (in 2D) or 8 (in 3D) equal rectangular boxes. Then, if possible, we crop each rectangular box to its minimal size still containing all points belonging to that box.
These ``divide'' and ``crop'' steps are illustrated in Figure \ref{fig: re-meshing procedure}.
This procedure is then recursively repeated for all boxes in which the variation exceeds the threshold, until a maximum level of recursion, $l$, is reached.
Once the recursion halts, we find the resulting reduced discretisation by representing each box by a single point at its centre.
To ensure a proper definition of the boundary, after recursive re-meshing, we add an evenly-spaced set of boundary points on the surface of the outermost box.

The recursive re-meshing method has two parameters that determine the resulting discretisation, the maximum variation within a box, $r_\text{max}$, and the maximum recursion level $l$.
The former dictates the smoothness of the tracer function, while the latter limits the overall recursion level, thereby limiting the smallest resolvable length scale. Increasing $r_\text{max}$ will result in a smoother representation of the tracer function, while increasing $l$ will result in better-resolved small-scale details.
We have implemented this recursive re-meshing algorithm efficiently in \textsc{Python} in \textsc{Magritte}.

\begin{table}
	\centering
	\caption{Comparing the number of points, time it takes to create a reduction, and time it takes to perfom a radiative transfer simulation.}\label{table: remesh_computation_time}
	\begin{tabular}{Sl | Sc  Sc  Sc}
		  & $N_{\text{points}}$ & $t_{\text{reduction}}$ [s] & $t_{\text{RT}}$ [s] \\
		\hline
		Original    & $1.1 \cdot 10^{6}$ &                   $-$ & $5.2\cdot 10^3$\\
        \hline 
		Reduced old & $9.2 \cdot 10^{4}$ & $2.2 \cdot 10^{2}$ & $1.9\cdot 10^2$ \\
		Reduced new & $8.5 \cdot 10^{4}$ & $6.0 \cdot 10^{0}$ & $2.3\cdot 10^2$
	\end{tabular}
\end{table}

\begin{figure*}
	\centering
	\includegraphics[width=\linewidth]{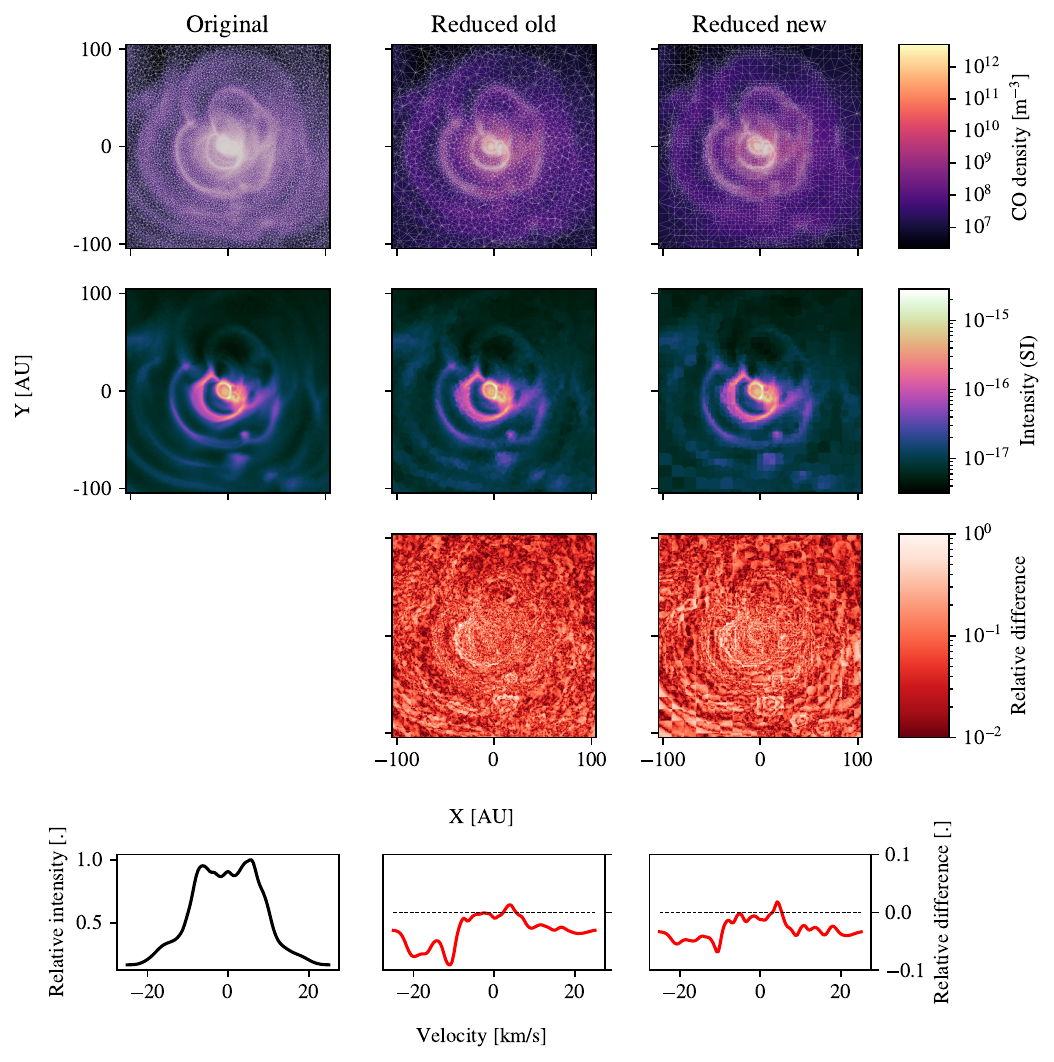}
	\caption{\textit{(Top row.)} Slice through the centre of the companion-perturbed stellar wind model showing the CO number density (cfr.\ Figure \ref{fig: AGB model density}), but now with a Delaunay tetrahedralisation of the spatial discretisation superimposed. \textit{(Second row.)} Synthetic observations of the CO $J=2-1$ line, showing only $v=0$ km/s, created with \textsc{Magritte}, using the different spatial discretisations.  \textit{(Third row.)} The relative difference in the synthetic observations, compared to the original model. \textit{(Bottom row.)} Integrated line profile obtained from the original model, and the relative differences for the reduced discretisations compared to the the original model.}\label{fig: combined image meshes and diffs}
\end{figure*}

\begin{figure}
	\includegraphics[width=\linewidth]{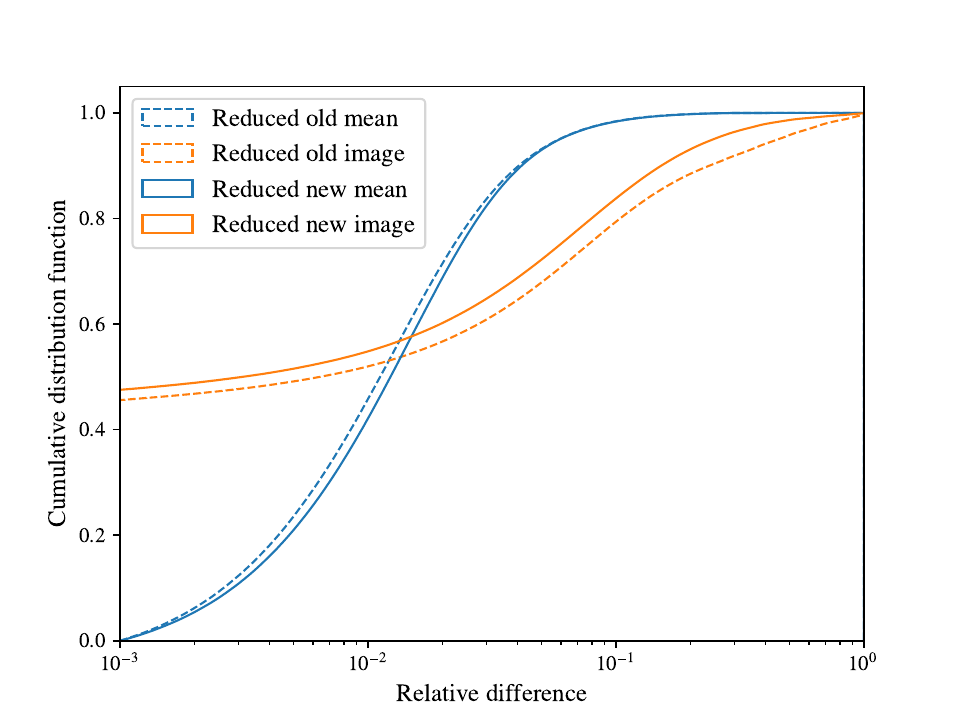}
	\caption{Cumulative distributions of the relative differences between results obtained on the reduced discretisations and the original model.
	}\label{fig: relative differences meshes mean line intensity}
\end{figure}

We can now demonstrate the utility of this new method by applying it to the SPH model of a companion-perturbed stellar wind, introduced in Section \ref{section: AGB binary}.
We chose the parameter values, $r_\text{max} = 0.4$ and $l=12$, to obtain similar resulting discretisations as the old re-meshing algorithm from \cite{deceuster_magritte_2020-1}, so we can compare the two.
The results are summarised in Table \ref{table: remesh_computation_time}.
Both methods reduce the number of points by about two orders of magnitude.
However, we can see that the new method is about 50 times faster.
We note that, even though the new method results in slightly fewer points, the time it takes to perform a radiative transfer simulation on it is slightly longer.
This can be understood, since the new method creates a slightly more regular discretisation, resulting on average in slightly more points per ray.
Nevertheless, radiative transfer simulations are still an order of magnitude faster compared to the original model. 
Figure \ref{fig: combined image meshes and diffs} shows the three different discretisations, the corresponding synthetic observations of the CO ($J=2-1$) line, and the relative differences of these with respect to the original model.
Qualitatively, both reduced discretisations trace the underlying tracer function equally well and both yield equally accurate synthetic observations.
To quantitatively asses the accuracy of the reduced discretisations, following \cite{deceuster_magritte_2020-1}, in Figure \ref{fig: relative differences meshes mean line intensity}, we plotted the cumulative distribution of relative differences between the radiation fields computed on the reduced discretisations and the original model.
Since for non-LTE line transfer the direction- and frequency-averaged mean intensity is important, we have plotted the relative differences for the mean intensity as well as for the resulting intensity on the images.
For both quantities, the relative differences are below 10\% for 80\% of the points in the model or pixels in the image.
The new method can thus produce equally good reduced discretisations at a fraction of the computational overhead.

\subsection{Adaptive Ng-acceleration}
\label{section: ng-acceleration}
In line radiative transfer, the emissivity and opacity of the medium depend on the radiation field in a non-trivial way.
To resolve this interdependence, the state of the medium and the radiation field are often alternately solved in an iterative way. 
The convergence of this iterative process can be notoriously slow, and, as a result, various techniques have been devised to accelerate it \citep[see e.g.][]{olson_rapidly_1986-1, rybicki_accelerated_1991-1}.
One of these techniques is Ng-acceleration \citep{ng_hypernetted_1974, olson_rapidly_1986-1}.
This is a general acceleration-of-convergence technique and is known under various other names and has various related variants in other application domains \citep[see e.g.][]{sidi_convergence_2020}.
It aims to predict the fixed-point of the iterative process by a linear combination of the $N_{\text{steps}}$ last iterates.
In the classical formulation, the value of $N_{\text{steps}}$ is a fixed predefined hyper-parameter, and it was also implemented like this in \textsc{Magritte} \citep{deceuster_magritte_2020}.

We noted, however, that this value can adaptively be chosen to obtain even better convergence.
We call this ``adaptive Ng-acceleration'', and we determine $N_{\text{steps}}$ as follows.
After each regular iteration, we compute the Ng-prediction, using the last $N$ regular iterates.\footnote{To be precise, we only include iterates \emph{after} the last Ng-prediction, if a previous Ng-prediction was already made.}
Then, we decide whether or not to use this prediction as the next iteration step.
The criterion depends on the convergence of the running Ng-predictions versus the convergence of the regular iterates. 
We only use the Ng-prediction as the next iterate if the maximum relative change in the last two running Ng-predictions is smaller than the maximum relative change in the last two regular iterates.
Stated differently, we only use the Ng-prediction as the next iterate if the Ng-predictions converge faster than the regular iterates.
In this way, we keep on using more previous iterates in the Ng-prediction, i.e.\ $N$ keeps on growing, until we eventually use the Ng-prediction.
If the criterion remains unsatisfied for many iterations, of course, we cannot keep on storing the previous iterates indefinitely.
Therefore, we also introduce a maximum, $N_{\max}$, after which we use the Ng-prediction as new iterate anyway.

This adaptive version of Ng-acceleration requires a way to conveniently compute the Ng-prediction for any number $N$ of previous iterations.
We present only the resulting equations here and provide a full derivation in \ref{Appendix: General Ng-acceleration}.
The Ng-prediction for the fixed-point of an iterative process is a linear combination of the $N$ last iterates,
\begin{align}\label{eq: fixed point linearity}
\boldsymbol{n}_{\star}
=
\sum_{n=1}^{N} c_n \, \boldsymbol{n}_{-n} ,
\end{align}
where we used negative indices to denote elements, starting from the last one counting backwards (like in \textsc{Python}).
To ensure normalisation of the prediction, which is important e.g.\ for level populations, we require the coefficients, $c_{n}$, to sum to one.
Defining the residuals, $\boldsymbol{r}_{n} \equiv \boldsymbol{n}_{-n}-\boldsymbol{n}_{-n+1}$, and accumulating them in a matrix,
\begin{equation}
    R^{\text{T}}
    \equiv
    \begin{bmatrix}
    \boldsymbol{r}_1&
    \boldsymbol{r}_2&
    \dots&
    \boldsymbol{r}_{N-1}
\end{bmatrix}
\end{equation}
the resulting coefficient vector, $\boldsymbol{c}$, for the Ng-prediction reads,
\begin{align}
\boldsymbol{c}
\ = \
\frac{\big(R R^{\text{T}}\big)^{-1}\boldsymbol{1}}{\boldsymbol{1}^T\big(R R^{\text{T}}\big)^{-1}\boldsymbol{1}} ,
\end{align}
in which $\boldsymbol{1}$ denotes a $(N-1)$-dimensional vector of ones.
Note that the matrix that needs to be inverted is $(N-1 \times N-1)$-dimensional and does not depend on the dimension of the vector iterates, $\boldsymbol{n}$.
Since we expect to require only a relatively small number of iterates, we expect the matrix inversion to be cheap.

\begin{figure}
	\includegraphics[width=\linewidth]{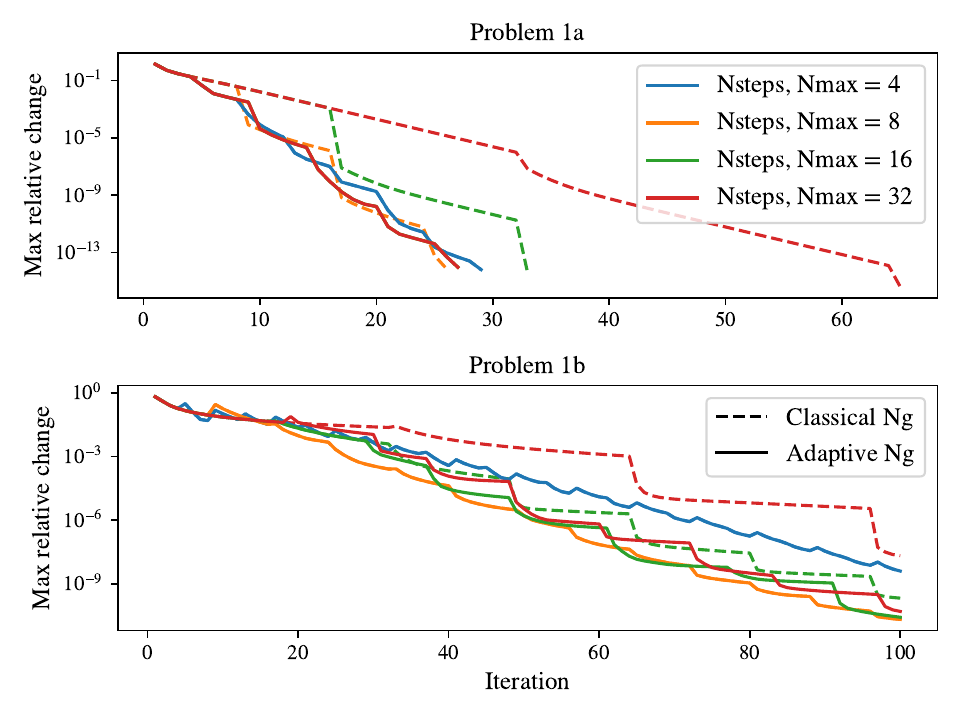}
	\caption{Convergence behavior of the classical and adaptive Ng-acceleration method  applied to the \cite{van_zadelhoff_numerical_2002} problem 1a/b, using different values for $N_{\text{steps}}$ and $N_{\text{max}}$. Both legends are for both plots.}
	\label{fig: ng-accel vanzadelhoff}
\end{figure}

We can demonstrate the performance of our adaptive Ng-acceleration scheme on the non-LTE benchmark problems by \cite{van_zadelhoff_numerical_2002}.
We consider what they call ``problem 1a/b'', which describes non-LTE line transfer in a spherically-symmetric steady-state model for a fictitious two-level species.
The a/b refers to the density of the fictitious species, resulting in a low/high optical depth medium.
Hence, we expect convergence for problem 1a, with lower optical depth, to be faster than  for 1b.
Figure \ref{fig: ng-accel vanzadelhoff} shows the convergence behaviour of classical and adaptive Ng-acceleration on this benchmark problem, for several choices of $N_{\text{steps}}$, in the classical, and $N_{\max}$, in the adaptive algorithm.
Adaptive Ng-acceleration consistently shows faster convergence.
Also when $N_{\max}$ is chosen relatively large (e.g.\ 16 or 32), the algorithm timely injects the Ng-prediction for optimal convergence.
As a result, for large enough $N_{\max}$, it does not affect the convergence behaviour anymore.
The choice of $N_{\max}$ should therefore only be informed by the amount of memory available.

\subsection{Semi-analytical approximation for line optical depths}
\label{section: optical depth integrating}
Optical depth plays a crucial role in radiative transfer, as it defines the relevant length-scales.
In a spatial discretisation, the optical depth between two neighbouring points, $\boldsymbol{x}_{0}$ and $\boldsymbol{x}_{1}$, is defined in terms of the opacity, $\chi$, as,
\begin{align}\label{eq: optical depth}
    \Delta\tau
    =
    \Delta x \int_{0}^{1} \text{d}\lambda \ \chi(\lambda) ,
\end{align}
where, $\lambda \in [0, 1]$, is a parameter interpolating between $\boldsymbol{x}_{0}$ at $\lambda=0$, and $\boldsymbol{x}_{1}$ at $\lambda=1$, and $\Delta x \equiv \left|\left| \boldsymbol{x}_{1}-\boldsymbol{x}_{0}\right|\right|$, is the Euclidean distance between the two points.

\begin{figure}
    \centering
	\includegraphics[width=0.9\linewidth]{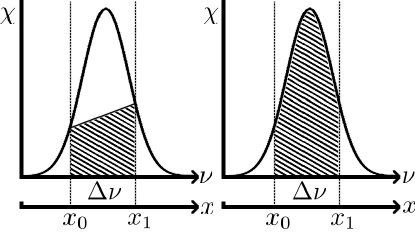}
    \caption{Illustration of the effect of a velocity gradient on line optical depth. \textit{(Left.)} The resulting integral assuming a linear integrant. \textit{(Right.)} The resulting integral by analytic integration of the Gaussian factor. }
    \label{fig: optical depth computation differences}
\end{figure}

In previous versions of \textsc{Magritte} \citep{deceuster_magritte_2020-1}, the integral in the optical depth equation (\ref{eq: optical depth}) was approximated using the trapezoidal rule,
\begin{align}\label{eq: trapezoidal rule}
    \Delta \tau = \Delta x\left(\frac{\chi_0+\chi_{1}}{2}\right),
\end{align}
which implicitly assumes that the opacity can be well-approximated by a linear function between the two points.
The subscripts indicate where the function is evaluated, i.e.\ $\chi_{i} \equiv \chi(\boldsymbol{x}_{i})$.
As long as the opacity is well-sampled by the spatial discretisation, this is a valid assumption.
However, this becomes difficult for line opacities when large velocity gradients are involved.
The line opacity of a transition $i \leftrightarrow j$, can be written as,
\begin{align}
\chi = \chi^{ij} \, \phi^{ij}(\nu) ,
\end{align}
where the first factor only depends on the position in the medium and the second factor is the line profile function that carries all frequency dependence.
In \textsc{Magritte}, we assume the line profile to be dominated by Doppler shifts caused by thermal and turbulent motions in the medium.
This results in a Gaussian line profile,
\begin{align}
\phi^{ij}(\nu) = \frac{1}{\delta\nu_{ij}\sqrt{\pi}} \, \exp\left[- \left(\frac{\nu-\nu_{ij}}{\delta\nu_{ij}}\right)^{2}\right] ,
\label{eq: gaussian}
\end{align}
where $\nu_{ij}$ is the line centre, and the line width is defined as,
\begin{align}\label{eq: line width}
\delta\nu_{ij} = \frac{\nu_{ij}}{c}\sqrt{\frac{2k_\text{B}T}{m_{\text{spec}}} + v_{\text{turb}}^2} ,
\end{align}
with $k_\text{B}$ Boltzmann's constant, $T$ the local gas temperature, $m_{\text{spec}}$ the molecular mass of the gas, and $v_{\text{turb}}$ the turbulent velocity.
The line profile is typically very narrow in frequency space and, therefore, very sensitive to velocity gradients in the medium due to Doppler shifts.
For the optical depth increment (\ref{eq: optical depth}), computed at a particular frequency, $\nu$, this implies,
\begin{align}\label{eq: optical depth 2}
    \Delta\tau (\nu)
    =
    \Delta x \int_{0}^{1} \text{d}\lambda \ \chi^{ij}(\lambda) \, \phi^{ij}\big(\nu+\lambda\Delta\nu\big) ,
\end{align}
where $\Delta \nu$ is the Doppler shift, caused by the velocity difference, $\Delta \boldsymbol{\mathrm{v}}$, between $\boldsymbol{x}_{0}$ and $\boldsymbol{x}_{1}$, projected along a direction $\boldsymbol{\hat{n}}$,
\begin{align}
    \Delta \nu  = \frac{\boldsymbol{\hat{n}} \cdot \Delta \boldsymbol{\mathrm{v}}}{c} \, \nu.
\end{align}
The problem is that the integrant in (\ref{eq: optical depth 2}) is not necessarily well-approximated by a linear function, as illustrated in Figure \ref{fig: optical depth computation differences}.
In previous versions of \textsc{Magritte}, this was resolved by subdividing the interval between neighbouring points if the Doppler shift was too large, such that on each subdivision the integrant was approximately linear \citep{deceuster_magritte_2020-1}.
Although effective, when large velocity gradients are present, many subdivisions will be necessary, causing significant computational overhead.
Therefore, in the current version of \textsc{Magritte}, we provide a semi-analytical approximation for the non-linear part of the integrant.
We do this by approximating $\chi_{ij}$ by its average between the points, and analytically integrating the remaining Gaussian factor,
\begin{align}\label{eq: main text single line optical depth}
    \Delta\tau(\nu)
    &=
    \Delta x \left(\frac{\chi^{ij}_0+\chi^{ij}_1}{2}\right) \int_{0}^{1} \text{d}\lambda \ \phi^{ij}\big(\nu+\lambda\Delta\nu\big) , \\
    &=
    \Delta x
    \left(\frac{\chi^{ij}_0+\chi^{ij}_1}{2}\right)\left(\frac{\text{Erf}\left(\frac{\nu_{ij}+\Delta\nu-\nu}{\delta\nu_{ij}}\right)-\text{Erf}\left(\frac{\nu_{ij}-\nu}{\delta\nu_{ij}}\right)}{2\Delta\nu}\right).
    \label{eq:semi-analytic}
\end{align}
In principle, the line width is also position-dependent but we assume that it can be well-approximated using the average local gas temperature and turbulent velocity between the two points.
Equation (\ref{eq:semi-analytic}) allows us to accurately account for Doppler shifts of the line profile at a small and fixed computational cost.

\begin{figure}
	\centering
    \includegraphics[width=\linewidth]{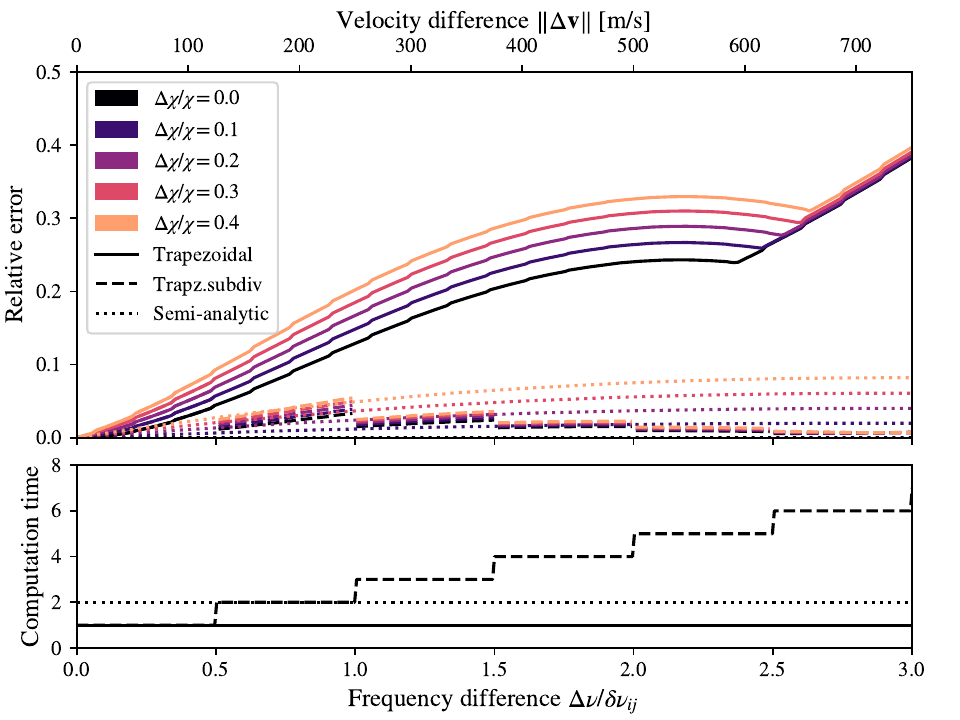}
	\caption{Relative error and computational cost of computing line optical depths, comparing the trapezoidal rule, the trapezoidal rule with subdivision, and the semi-analytic method.
}\label{fig: optical depth integration}
\end{figure}

Figure \ref{fig: optical depth integration} shows the relative error and computational cost for the line optical depth computation using the trapezoidal rule, the trapezoidal rule with subdivision, and the semi-analytical method.
With the semi-analytic method, we achieve comparable accuracy as with the trapezoidal rule with subdivision, but at a lower computational cost, especially for large velocity gradients.
The semi-analytic method is about twice as slow as the trapezoidal rule (without subdivision).
Therefore, in \textsc{Magritte}, for small Doppler shifts, i.e.\ $\Delta \nu / \delta\nu_{ij} < 0.35$, we still use the trapezoidal rule, while for larger Doppler shifts, we revert to the semi-analytical method to compute line optical depths.

\section{Current limitations \& future work}
\label{sec:limitations}
\textsc{Magritte} currently only supports line radiative transfer without any continuum emission or absorption within the medium.\footnote{The boundary conditions can, of course, contain continuum radiation, and are, in fact, usually continuum sources (e.g.\ cosmic microwave background).}
Although this might be a safe assumption when modelling low-$J$ lines in stellar atmospheres, in the future we would like to be able to include dust emission and absorption, which would be important for line transfer in higher-$J$ lines
\citep[see e.g.][]{matsumoto_self-consistent_2023}.
This will be included in a future version.

\textsc{Magritte} also cannot model radiation scattering.
Efficiently modelling scattering would probably require a significant revision of the entire modelling strategy.
Therefore, we do not plan to include it, at least not in the near future.
Nevertheless, we have been investigating ways to artificially model the effects of scattering (Su et al.\ in prep.), but these are currently beyond the short-term development scope of \textsc{Magritte}.

In the initial design of \textsc{Magritte}, considerable efforts were made to be able to leverage distributed computing systems \citep[see e.g.][]{deceuster_magritte_2020}.
We noticed, however, that for our target applications, rather than to simulate a single large model, it is more important to be able to efficiently simulate large grids of medium-sized models \citep[see e.g.][]{coenegrachts_unusual_2023-1, vandesande_modelling_2024}.
These medium-sized models typically fit on a single compute node.
Moreover, our discretisation reduction techniques help us fit also larger models on a single node \citep[see e.g.][and Section \ref{section: remeshing}]{deceuster_magritte_2020-1}.
Therefore, we currently focus on single-node performance, and have, at least temporarily, discontinued support for simulating on distributed systems. 
Still, most of the required infrastructure is still in the code base, so this could be restored in the future.

Previous versions of \textsc{Magritte} also included an adaptive directional discretisation method that could trace more rays in directions of interest (e.g.\ with high column density), to obtain more accurate direction-averaged mean intensities \citep{deceuster_magritte_2020-1}.
In the current version of \textsc{Magritte}, however, this feature was removed, since the relatively small gain in accuracy did not outweigh the significant maintenance overhead that the rather cumbersome implementation imposed.
Nevertheless, we are working on a more elegant re-implementation of this idea, which will become available in a future version of \textsc{Magritte}.

\section{Conclusion}
\label{conclusion}
We have presented the first long-term stable release of \textsc{Magritte}, an open-source software library for line radiative transfer.
We have demonstrated its use with two applications creating synthetic spectral line observations, one for an analytic model of a protoplanetary disk and one for a numerical SPH model of a complex companion-perturbed stellar wind.
Further, we have introduced the overall design strategy for \textsc{Magritte}, including the class structure, the application/programmer interface, and the code development and maintenance strategy, including automated testing.
Finally, we presented three key improvements over previous versions.
First, we have presented a simple and fast recursive re-meshing algorithm to efficiently create spatial discretisations of a model that can be tailored for radiative transfer simulations.
We have demonstrated that our new algorithm can provide similar quality spatial discretisations as before, but it is 50 times faster.
Second, we have presented an adaptive variation on the popular Ng-acceleration method to accelerate convergence in non-LTE line transfer simulations.
We have shown that an optimal value for the only hyper-parameter in the classical algorithm ($N_{\text{steps}}$) can be determined on-the-fly to consistently obtain optimal convergence.
Third, we have derived a semi-analytical approximation for the line optical depth,
which can accurately handle even large velocity gradients at a small and fixed computational cost.
This obviates the need for computationally more expensive sub-stepping methods.
In summary, \textsc{Magritte} is now a mature project with a significant application domain in astrophysical line radiation transport.

\section*{Acknowledgments}
We thank Jolien Malfait for kindly providing the SPH model of the companion-perturbed stellar wind. We thank the anonymous reviewers for their helpful comments.
TC is a PhD Fellow of the Research Council - Flanders (FWO), grant 1166722N. FDC is a Postdoctoral Research Fellow of the Research Foundation - Flanders (FWO), grant 1253223N. LD acknowledges support from the KU Leuven C1 MAESTRO grant C16/17/007, the FWO research grant G099720N and the KU Leuven C1 BRAVE grant C16/23/009.

\section*{Data Availability}\label{section: data availability}

\textsc{Magritte} is an open-source software library publicly available at \url{github.com/Magritte-code/Magritte}.
The benchmarks presented in this paper are part of the source code.
A fork of the source code, including \textsc{Jupyter} notebooks and \textsc{Python} scripts used for the simulations and creating the plots in this paper, can be found at \url{github.com/Magritte-code/Magritte_Paper_Astronomy_Computing}.

\bibliographystyle{elsarticle-harv} 
\bibliography{Library_2024_06_18} 

\appendix

\section{Derivation of Ng-acceleration}
\label{Appendix: General Ng-acceleration}
The adaptive Ng-acceleration scheme presented in Section \ref{section: ng-acceleration} requires a way to conveniently predict the Ng-acceleration step, $\boldsymbol{n}_{\star}$, as a linear combination of the $N$ last iterates,
\begin{align}
\boldsymbol{n}_{\star}
=
\sum_{n=1}^{N} c_n \, \boldsymbol{n}_{-n} ,
\label{eq:lin}
\end{align}
where we used negative indices to denote elements, starting from the last one counting backwards (like in \textsc{Python}).
To ensure normalisation of the prediction, we require the coefficients, $c_{n}$, to sum to one, i.e.\ $\boldsymbol{c}^{\text{T}}\boldsymbol{1}=1$, where $\boldsymbol{1}$ is a conforming vector of ones.
The goal is now to find the coefficients, $c_{n}$, such that $\boldsymbol{n}_{\star}$ is the ``optimal prediction'' in some sense for the fixed-point of the iterative process that we are interested in.
Before we can find these, we first need to specify what we mean by ``optimal prediction''.
We can formalise the iterative process by an operator, $\Lambda$, acting on our iterates, $\boldsymbol{n}$.
For non-LTE line transfer, for instance, the $\Lambda$-operator would comprise the computation of the radiation field and the solution of the statistical equilibrium equations, while $\boldsymbol{n}$ would be the level populations \citep[see][for the particular implementation in \textsc{Magritte}]{deceuster_magritte_2020}.
The true fixed-point, $\boldsymbol{n}_{*}$, of this iterative process would then be defined by,
\begin{equation}
    \boldsymbol{n}_{*} = \Lambda[\boldsymbol{n}_{*}] .
    \label{eq:fixed-point}
\end{equation}
In non-LTE line transfer, this fixed-point would correspond to the level populations that are self-consistent with the radiation field.
Using this formalism, we can define the optimal coefficients, $c_{n}$, as the solution of the following constraint minimisation problem,
\begin{align}
\boldsymbol{c}
=
\underset{\boldsymbol{c}: \, \boldsymbol{c}^{\text{T}}\boldsymbol{1}=1}{\text{argmin}}
\left\{
    \big\|\boldsymbol{n}_{\star}-\Lambda[\boldsymbol{n}_{\star}]\big\|^2
\right\}
\label{eq:sense}
\end{align}
In words, we want the fixed-point equation (\ref{eq:fixed-point}) to be satisfied as closely as possible, given the constraint $\boldsymbol{c}^{\text{T}}\boldsymbol{1}=1$.
In order to solve this problem, we need to assume that the $\Lambda$-operator is linear.
Although, in general, this assumption will not hold, it will hold approximately near the fixed-point, which is the region we are interested in anyway.
Assuming linearity of the $\Lambda$-operator and substituting equation (\ref{eq:lin}), yields,
\begin{align}
\boldsymbol{c}
&=
\underset{\boldsymbol{c}: \, \boldsymbol{c}^{\text{T}}\boldsymbol{1}=1}{\text{argmin}}
\left\{
    \left\|\sum_{n=1}^{N} c_n \, \boldsymbol{n}_{-n} - \sum_{n=1}^{N} c_n \, \Lambda[\boldsymbol{n}_{-n}]\right\|^2
\right\} , \\
&=
\underset{\boldsymbol{c}: \, \boldsymbol{c}^{\text{T}}\boldsymbol{1}=1}{\text{argmin}}
\left\{
    \left\|\sum_{n=1}^{N} c_n  \big( \boldsymbol{n}_{-n} - \boldsymbol{n}_{-n+1} \big) \right\|^2
\right\} ,
\end{align}
where, in the second line, we used that $\boldsymbol{n}_{i+1} = \Lambda[\boldsymbol{n}_{i}]$.
Defining the residuals, $\boldsymbol{r}_{n} \equiv \boldsymbol{n}_{-n}-\boldsymbol{n}_{-n+1}$, and accumulating them in a matrix,
\begin{equation}
    R^{\text{T}}
    \equiv
    \begin{bmatrix}
    \boldsymbol{r}_1&
    \boldsymbol{r}_2&
    \dots&
    \boldsymbol{r}_{N-1}
\end{bmatrix}
\end{equation}
the constraint minimisation problem can be written as,
\begin{align}
\boldsymbol{c}
&=
\underset{\boldsymbol{c}: \, \boldsymbol{c}^{\text{T}}\boldsymbol{1}=1}{\text{argmin}}
\left\{
    \boldsymbol{c}^{\text{T}} \, R R^{\text{T}} \, \boldsymbol{c}
\right\} , \\
&=
\underset{\boldsymbol{c} , \, \lambda}{\text{argmin}}
\left\{
    \boldsymbol{c}^{\text{T}} \, R R^{\text{T}} \, \boldsymbol{c} \ + \ \lambda \left(\boldsymbol{c}^{\text{T}}\boldsymbol{1} - 1\right)
\right\} , 
\end{align}
where, in the second line, we accounted for the constraint by introducing a Lagrange multiplier, $\lambda$.
At the minimum, both the gradient with respect to $\boldsymbol{c}$ and the gradient with respect to $\lambda$ should vanish, yielding equations,
\begin{align}
0 &= 2\boldsymbol{c}^{\text{T}} R R^{\text{T}} \ + \ \lambda\boldsymbol{1}^{\text{T}} \\
0 &= \boldsymbol{c}^{\text{T}}\boldsymbol{1} - 1
\end{align}
Solving for $\lambda$, yields,
\begin{align}
\lambda
=
\frac{-2}{\boldsymbol{1}^{\text{T}} \big(R R^{\text{T}}\big)^{-1}\boldsymbol{1}} ,
\end{align}
while solving for $\boldsymbol{c}$ and substituting the result for $\lambda$ yields,
\begin{align}
\boldsymbol{c}
\ = \
\frac{\big(R R^{\text{T}}\big)^{-1}\boldsymbol{1}}{\boldsymbol{1}^T\big(R R^{\text{T}}\big)^{-1}\boldsymbol{1}} .
\end{align}
These are the coefficients that make equation (\ref{eq:lin}) optimally predict the fixed-point of the iterative process defined by $\Lambda$, in the sense of equation (\ref{eq:sense}), assuming linearity.

	
	
	
	

	
\end{document}